\documentclass[aps,prx,twocolumn,nofootinbib,notitlepage]{revtex4-1}
\usepackage{graphicx}
\usepackage{amssymb}
\usepackage{hyperref}
\usepackage{amsmath}
\usepackage{amsfonts}
\usepackage{subfigure}
\usepackage{dsfont}
\usepackage{times}
\usepackage{dcolumn}
\usepackage{esint}
\usepackage{mathrsfs}
\usepackage{bm}
\usepackage{multirow}
\usepackage{color}
\usepackage{datetime}
\usepackage{braket}

\begin{document}
\title{Entanglement Signature of   Hinge Arcs, Fermi Arcs, and Crystalline Symmetry Protection in Higher-Order Weyl Semimetals}

\author{Yao Zhou}
\affiliation{School of Physics, State Key Laboratory of Optoelectronic Materials and Technologies, and Guangdong Provincial Key Laboratory of Magnetoelectric Physics and Devices, Sun Yat-sen University, Guangzhou, 510275, China}
\author{Peng Ye}
\email{yepeng5@mail.sysu.edu.cn}
\affiliation{School of Physics, State Key Laboratory of Optoelectronic Materials and Technologies, and Guangdong Provincial Key Laboratory of Magnetoelectric Physics and Devices, Sun Yat-sen University, Guangzhou, 510275, China}

\date{\today}

\begin{abstract}
 The existence of $1/2$ modes in the entanglement spectrum  (ES) has been shown to be a powerful quantum-informative signature of     boundary  states of  gapped topological phases of matter, e.g., topological insulators and topological superconductors, where the finite bulk gap allows us to establish a crystal-clear correspondence between   $1/2$ modes and boundary states. Here we investigate the recently proposed higher-order Weyl semimetals (HOWSM), where   bulk supports gapless higher-order Weyl nodes and    boundary supports hinge arcs and Fermi arcs. We find that the aim of unambiguously identifying higher-order boundary states ultimately drives us to make full use of eigen quantities of the entanglement Hamiltonian:   ES as well as Schmidt vectors (entanglement wavefunctions, abbr. EWF). We demonstrate that, while both hinge arcs and Fermi arcs contribute to $1/2$ modes,  the EWFs corresponding to hinge arcs and Fermi arcs are respectively localized on the virtual hinges and surfaces of the partition.   Besides, by means of  various symmetry-breaking partitions, we can identify the minimal crystalline symmetries that protect  boundary states. Therefore, for gapless topological phases such as HOWSMs, we can combine  ES and EWF to universally identify   boundary states and   potential symmetry requirement.  While   HOWSMs are prototypical examples of gapless phases, our work sheds light on general theory of entanglement signature in gapless topological phases of matter.
\end{abstract}

\maketitle

{\color{blue}\emph{Introduction.}}---
Boundary states~\cite{hasanColloquium2010,bansilColloquium2016,qiGeneral2006,wenColloquium2017}  are  a significant feature of topological phases of matter, which  realizes the scenario of bulk-boundary holographic correspondence in condensed matter systems \cite{hartnollQuantum2021}. To characterize this correspondence, from the quantum-informative perspective, topological  entanglement entropy (TEE)~\cite{kitaevTopological2006,levinDetecting2006,amicoEntanglement2008}   was proposed, which encodes the total quantum dimension of anyon contents. While nonzero TEE indicates nontrivial  topological order in the bulk,  it was also re-investigated and generalized into the field of fracton topological orders~\cite{williamsonSpurious2019,nandkishoreFractons2019,pretkoFracton2020} as well as excited states~\cite{wenEntanglement2018}. Furthermore, Li and Haldane~\cite{liEntanglement2008} employed, instead of TEE such a single number,  the full spectrum (i.e., entanglement spectrum, abbr. ES) of the reduced density matrix to characterize fractional quantum Hall states, and numerically discovered the relation between its ES and edge mode spectrum on physical boundaries. A   theoretical investigation into this  relation was later given by  Ref.~\cite{qiGeneral2012}. Ref.~\cite{pollmannEntanglement2010} found that two degenerate entanglement modes (EMs)  are elegantly related to the doubly degenerate  physical edge spectrum in  Haldane chains, which together with Ref.~\cite{guTensorentanglementfiltering2009} triggered tremendous advances on  symmetry-protected topological phases (SPT) in strongly correlated bosonic/spin systems~\cite{chenLocal2010,chenSymmetryProtected2012}.

In  free-fermion systems, entanglement Hamiltonian is quadratic form~\cite{peschelReduced2009} and  its ES  is fully determined by   the single-particle Green's function~\cite{chungDensitymatrix2001,peschelCalculationreduceddensity2003,peschelReduced2009}. Unfortunately, since entanglement Hamiltonian  is usually not a sparse matrix, the ES in most cases can only be calculated numerically~\cite{peschelReduced2009,peschelReduced2004,cheongOperatorbased2004} despite the non-interacting nature of free-fermion systems. But ES of free-fermion systems still possesses nice properties on the aspects of duality transformations and block Toeplitz matrices~\cite{leePositionmomentum2014,leeFreefermion2015}, which has recently been generalized to non-Hermitian systems~\cite{chenEntanglement2020,chenQuantum2022}. Meanwhile, by flattening the Hamiltonian, Fidkowski~\cite{fidkowskiEntanglement2010} found an exact result that $1/2$ EMs are originated from topological edge states.  Recently, the study of free-fermion topological phases has been generalized to  the higher-order topological insulators~\cite{zhangSurface2013,benalcazarQuantized2017,langbehnReflectionSymmetric2017,songEnsuremath22017,schindlerHigherorder2018} in which   topologically protected corner and hinge states are located on boundary of the bulk phase. In these systems, Refs.~\cite{schindlerHigherorder,fukuiEntanglement2018} found that   the nested ES and the entanglement polarization can be used to characterize higher-order topology. Moreover, a relation between ES and the corner-induced filling anomaly in higher-order topological insulators was  discussed in Ref.~\cite{zhuIdentifying2020}. 

On the other hand, gapless topological phases of matter have also been widely studied in free fermion systems, such as   Weyl~\cite{wanTopological2011}, Dirac~\cite{wangDirac2012,youngDirac2012} and nodel-loop semimetals~\cite{burkovTopological2011,fangTopological2016}. It is thus natural  to ask how to use quantum entanglement to clearly identify topological boundary states in gapless systems.  This question is challenging since the existence of bulk gap is very crucial for establishing theoretical arguments on the correspondence between ES and physical boundary spectrum in the previous studies of gapped systems~\cite{qiGeneral2012,fidkowskiEntanglement2010}.  Furthermore, this situation becomes even worse in recently proposed  systems where boundary spectrum  incorporates very complex structure, such as higher-order Dirac (HODSM)~\cite{linTopological2018}, higher-order Weyl (HOWSM)~\cite{ghorashiHigherOrder2020,wangHigherOrder2020,weiHigherorder2021} and higher-order nodal-loop semimetals~\cite{wangBoundary2020}. For example, in HOWSMs~\cite{wangHigherOrder2020,ghorashiHigherOrder2020,weiHigherorder2021},   there are two types of Weyl nodes in the bulk: the conventional Weyl nodes are projected onto boundary, which forms  surface Fermi arcs; the 2nd-order Weyl nodes are attached by topological hinge arcs locating on the hinges of HOWSMs. Thus, it is interesting to ask the following more specific question: how to crystal-clearly characterize boundary states of gapless topological phases exemplified by HOWSMs in a quantum-informative way?

This work is  motivated to address this issue. We propose to make full use of eigen quantities of the entanglement Hamiltonian:   ES as well as Schmidt vectors (entanglement wavefunctions, abbr. EWF). In this way,  we are allowed to clearly identify all boundary states of HOWSMs.  More specifically,  while both hinge arcs and Fermi arcs contribute to $1/2$ EMs,  the EWFs corresponding to hinge arcs and Fermi arcs are respectively localized on the virtual hinges and surfaces of the partition. Therefore, in HOWSM systems, ES alone is  far more insufficient and potentially misleading in unambiguously identifying boundary states. Furthermore, by exhausting prototypical  symmetry-breaking partitions, ES and EWF as a whole can be applied to clearly identify the crystalline symmetries that protect boundary states of HOWSMs.  Our work   demonstrates the usefulness and necessity of EWFs in characterizing gapless topological phases of matter, where the EWF as an entanglement signature for topological system has been rarely employed in the literature to the best of our knowledge. As HOWSMs are just the tip of the iceberg of highly unexplored gapless topological phases, we expect the combination of ES and EWFs will play a critical role of entanglement signature of gapless topological phases in both free-fermion and strongly correlated systems.

{\color{blue}\emph{Symmetry of ES and EWF from correlation}}---
Recently, HOWSMs are introduced in different models~\cite{ghorashiHigherOrder2020,wangHigherOrder2020,weiHigherorder2021} with different crystalline symmetries. Here we  start with the model in Ref.~\cite{ghorashiHigherOrder2020}, which is realized by adding a symmetry-breaking term into a HODSM model. To be specific, this HOWSM model is given by $H_{\text{HOWSM}}(\bm{k})=H_{\text{HODSM}}(\bm{k})+i m \Gamma_{1}\Gamma_{3}$, where $H_{\text{HODSM}}(\bm{k})= (f(z)+\cos k_{x})\Gamma_{4} +(f(z)+\cos k_{y})\Gamma_{2}+\sin k_{x}\Gamma_{3}+\sin k_{y}\Gamma_{1}$ with $f(z)=\gamma+1/2\cos k_{z}$. Here,  Gamma matrices are  defined as $\Gamma_{j}=-\sigma_{j}\otimes\tau_{2}$ $(j=1,2,3)$, $\Gamma_{4}=\mathbb{I}\otimes\tau_{1}$ and $\Gamma_{5}=\mathbb{I}\otimes\tau_{3}$, where $\sigma_{i}$ and $\tau_{i}$ ($i=1,2,3$) are two independent sets of Pauli matrices.    $H_{\text{HOWSM}}(\bm{k})$ respects four-fold rotation symmetry $C^{z}_{4}$, deformed mirror symmetries $M_{x}T$, $M_{y}T$ ($T$ is time-reversal) and inversion symmetry $I$. For traditional method characterizing the topological boundary states of HOWSM system, we should adopt two topological invariants, i.e., Chern number and quadrupole moment~\cite{wangHigherOrder2020,ghorashiHigherOrder2020}. However, the method of unified characterizing topological boundary states for HOWSM system is absent.

Since we aim to apply entanglement to give a unified description for the topological boundary states of HOWSM system,  it is useful to start with preliminaries about density matrix $\rho=\ket{G}\bra{G}$, where $\ket{G}$ is the ground state of the system. When partitioning the system into two parts $A$ and $B$, and tracing over part $B$, then we obtain the reduced density matrix $\rho_{A}=\text{Tr}_{B}\ket{G}\bra{G}=\frac{1}{\mathcal{N}}\exp(-\mathcal{H}^{E})$, where the spectrum of entanglement Hamiltonian $\mathcal{H}^{E}$ is called entanglement spectrum. In free-fermion systems, the single-particle correlation matrix is defined as $C(\bm{r}_{i},\bm{r}_{j})=\bra{G}c^{\dagger}(\bm{r}_{i})c(\bm{r}_{j})\ket{G}$. When $\bm{r}_{i,j} \in A$, the correlation matrix can be written as $C^{A}(\bm{r}_{i},\bm{r}_{j})=\text{Tr}[\rho_{A}c^{\dagger}(\bm{r}_{i})c(\bm{r}_{j})]$. As discussed in Ref.~\cite{peschelCalculationreduceddensity2003}, one can simultaneously diagonalize $C^{A}(\bm{r}_{i},\bm{r}_{j})$ and $\mathcal{H}^{E}$, and their eigenvalues have an one-to-one correspondence relation $\varepsilon_{i}=\log[(\xi_{i})^{-1}-1] $, where $\varepsilon_{i}$ and $\xi_{i}$ are the eigenvalues of $\mathcal{H}^E$ and $C^A$ respectively. In this paper, we adopt $\xi_{i}$'s to represent ES. We note that, since the correlation matrix $C^{A}$ is not sparse matrix, then the entanglement Hamiltonian is  in fact a non-local Hamiltonian, leading to much computational cost in entanglement calculation.

Next, we give an efficient way to evaluate  the correlation matrix $C$. When a system has translational invariance, the Hamiltonian of the system can be rewritten as $H(\bm{k})$ in momentum space. Because $\bm{k}$ is a good quantum number, the correlation matrix $C(\bm{k})=\text{Tr}[\rho c^{\dagger}(\bm{k})c(\bm{k})]$ in momentum space is block-diagonal, and can be written explicitly as $ C(\bm{k})=[1+e^{H(\bm{k})}]^{-1}$. In terms of $C(\bm{k})$, the real-space correlation matrix can be rewritten as $C(\bm{r}_{i},\bm{r}_{j}) =\int_{BZ}\frac{d\bm{k}}{V_{BZ}}e^{- i\bm{k}\cdot(\bm{r}_{i}-\bm{r}_{j})}C(\bm{k}) $, where BZ is Brillouin zone. If $\bm{r}_{i,j}$ are restricted in  $A$, we immediately obtain the correlation matrix $C^{A}$ of the subsystem $A$. For the higher-order topological system, the topological phases are protected by the crystalline symmetries. Therefore, we propose the entanglement properties of higher-order phase are protected by crystalline symmetries, and should study how the crystalline symmetries affect the ES and EWF of HOWSMs. From the expression of correlation matrix, and to simplify the discussion, we shift the spectrum of correlation matrix along the $-y$ direction with $1/2$. Then the correlation matrix is rewritten as:
\begin{equation}
\!\tilde{C}^{A}(\bm{r}_{i},\bm{r}_{j}) \!=\!\int_{BZ}\frac{d\bm{k}}{V_{BZ}}e^{- i\bm{k}\cdot(\bm{r}_{i}-\bm{r}_{j})}\!\left[-\frac{1}{2}\tanh\frac{  H(\bm{k})}{2}\right].\!
  \label{eq_cor_matx_sym}
\end{equation}  
Here we use $\tilde{C}^{A}$ to study the symmetry of ES and EWF. From the above Eq.~\eqref{eq_cor_matx_sym}, the term $\tanh \frac{H}{2}$ in $\tilde{C}^{A}$ is determined by momentum-space Hamiltonian $H(\bm{k})$. Therefore, the symmetry of $H(\bm{k})$ would affect the ES and EWF. Meanwhile, as shown in Eq.~\eqref{eq_cor_matx_sym}, $\tilde{C}^{A}$ also depends on the lattice site $\bm{r}_{i,j}$ of the subsystem A, then the symmetry of ES and EWF also depends on the geometry of partition. Therefore, to study the symmetry of ES and EWF, we should consider the symmetry of $H(\bm{k})$ and classify the symmetry of partition. In general, the partition of a system would break the translational invariance. Then when the partition breaks the translational invariance of $n$ directions, we denote the partition as $n$-dimension partition ($n$D partition). Besides, the partition would also break the other crystalline symmetries, such as rotation and mirror symmetries etc., therefore we will  discuss this symmetry-breaking effect for ES and EWF in the following section.

\begin{figure}[htbp]
  \centering
  \includegraphics[width=7cm]{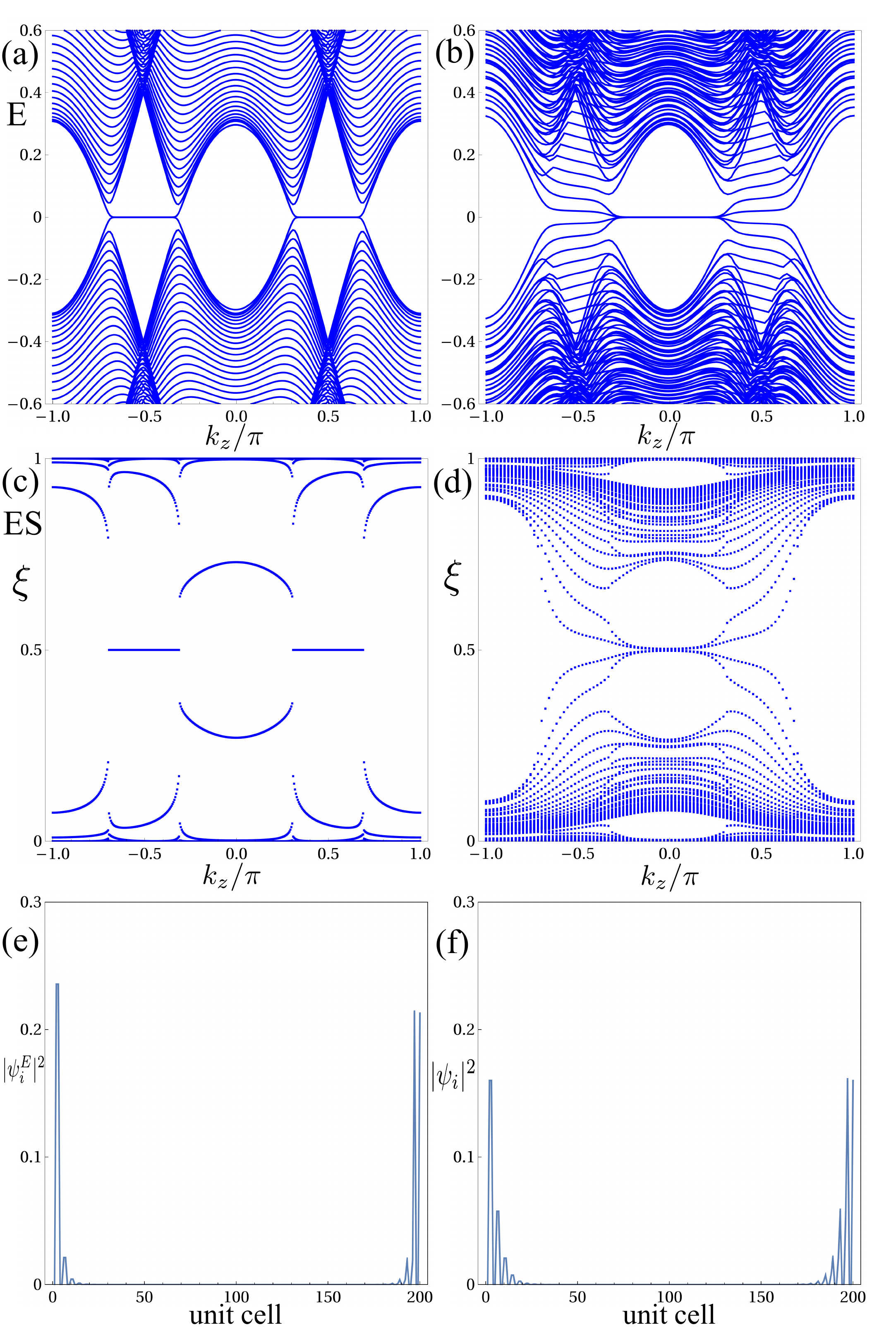}
  \caption{(a) and (b) are   the energy spectrum of HOWSMs as a function of $k_{z}$ with open boundary condition respectively along $x$ direction and along  $x$ and $y$ directions. (c) and (d) are   the ES of HOWSMs as a function of $k_{z}$ with $1$D and $2$D partitions respectively along $x$ direction  and along $x$ and $y$ directions. Here $k_{y}=0$ in (a) and (c). We choose a representative point $k_{z}=\pi/2$ to respectively demonstrate the amplitude distribution of EWF and physical wavefunction of Fermi arc in (e) and (f). These two kinds of wavefunctions are  localized on the virtual surface of partition and physical surface of system, respectively.  Here $\lambda=1$, $\gamma=-1$ and $m=0.4$.}\label{onedpart}
\end{figure}

{\color{blue}\emph{Entanglement signature of Fermi arcs}}---To characterize topological boundary states of HOWSMs, we first use a $1$D partition to study entanglement signature of Fermi arcs, in which case the partition breaks the translational invariance of one direction. Then, the remaining directions of this 3D system still have translational invariance, and correlation matrix is block-diagonal in the momentum space along the translational invariant directions. To simplify the discussion, we partition the HOWSM system along the $x$ direction, so $k_{y}$ and $k_{z}$ are good quantum numbers and the correlation matrix can be represented as $C(x, k_y ,k_z)$.   We numerically diagonalize $C(x, k_y ,k_z)$ to demonstrate ES in Fig.~\ref{onedpart}(c). We also obtain the physical energy spectrum with open boundary condition in $x$ direction,  as shown in Fig.~\ref{onedpart}(a). By comparing (a) and (c), we find that in the parameter regions with Fermi arcs, ES exactly supports $1/2$ EMs. Meanwhile, when the 3D HOWSM system is regarded as stacked $k_{z}$-dependent $2$D systems, the $k_{z}$ parameter regions with Fermi arcs have non-trivial Chern number. So we conclude that in HOWSM system, $1/2$ EM of ES is the significant entanglement signature of Fermi arcs.

Since both hinge arcs and Fermi arcs exist as topological boundary states of HOWSM,  ES alone is insufficient to identify Fermi arcs. For this purpose, we   consider the entanglement signature arising from EWFs. As discussed in the above section, we can diagonalize simultaneously the correlation matrix $C^{A}$ and the entanglement Hamiltonian $\mathcal{H}^{E}$ by using a unitary matrix. Then, we study the eigenfunction of correlation matrix for convenience. As illustrated in Fig.(\ref{onedpart})(e), we find that the EWF of a $1/2$ EM is localized in the virtual surfaces of the $1$D partition, which is similar to the physical wave function of Fermi arcs locating in the physical surface of HOWSM system as shown in Fig.~\ref{onedpart}(f). Therefore, we need to combine ES and EWFs together as entanglement signature for identifying Fermi arc. We numerically find this  signature is in fact model-independent (see Appendix \ref{app_secdhowsm}).

\begin{figure*}[htbp]
\centering
\includegraphics[width=13cm]{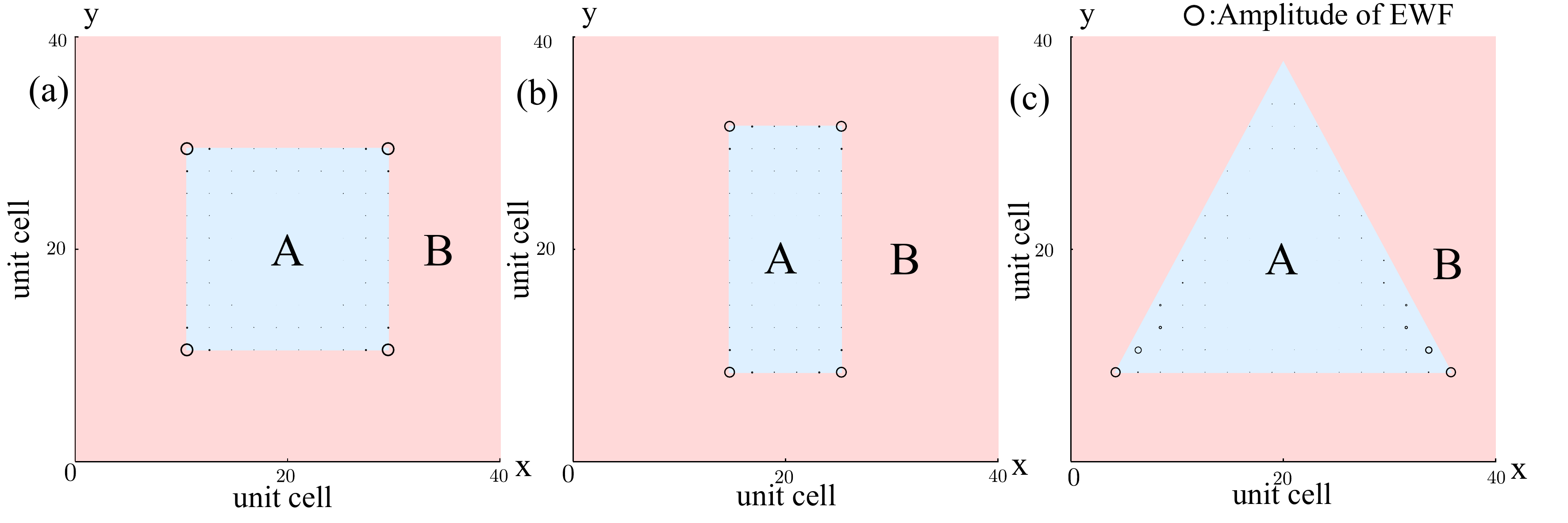}
\caption{In $z$ direction, these $2$D partitions have translational invariance, then we only plot $x$-$y$ plane to demonstrate these partitions. Here the corners of part-A is the virtual hinges of the partition.  (a) is the EWFs of $1/2$ EMs in part $A$ with square geometry preserving all crystalline symmetries of the HOWSM model. (b) is the  EWFs in part $A$ with rectangle geometry breaking rotation symmetry $C_{4}$ but preserving mirror symmetries $M_{y}T$ and $M_{x}T$. (c) is the EWFs in part $A$ with triangle geometry breaking rotation symmetry $C_{4}$ and mirror symmetry $M_{y}T$ but preserving mirror symmetry $M_{x}T$. }\label{twodpartwave}
\end{figure*}

\begin{figure}[htbp]
\centering
\includegraphics[width=7cm]{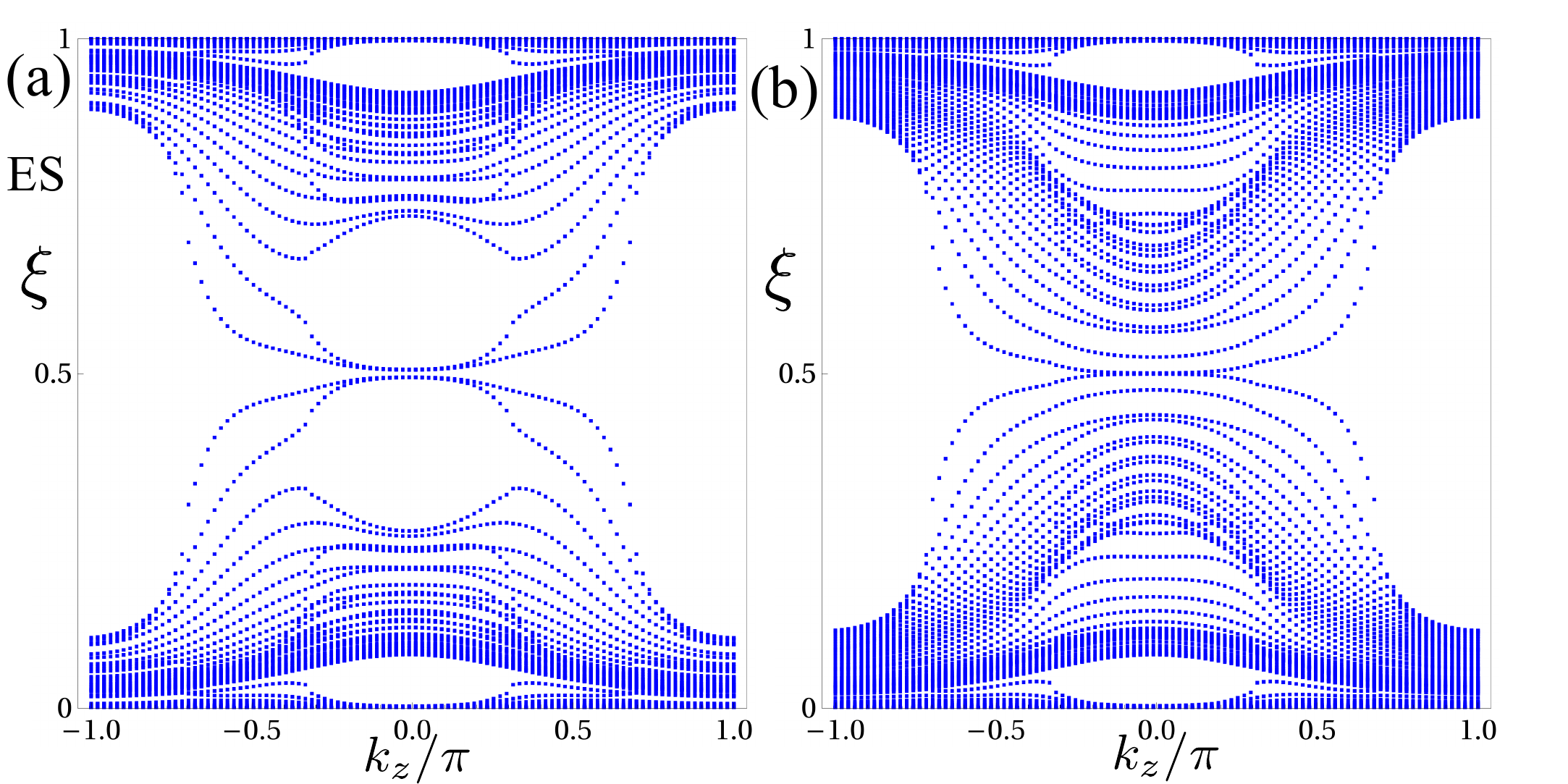}
\caption{(a) and (b) are respectively the ES related to the rectangle geometry in Fig.~\ref{twodpartwave} (b) and the triangle geometry in Fig.~\ref{twodpartwave} (c).  There are four-fold $1/2$ EMs and two-fold $1/2$ EMs in (a) and (b), respectively.}\label{twodpartbreak}
\end{figure}

{\color{blue}\emph{Entanglement signature of   hinge arcs}}---For the conventional $d$ dimension topological phases, topological boundary states often appear in the $d-1$ dimension boundary of the system. However, for the higher-order topological phase, topological boundary states would appear in the $d-n$ dimension boundary, where the integer $n>1$. Therefore, this phase is called   $n$th-order topological phase. In this paper, due to the topological boundary states of HOWSMs depending on the dimension of physical boundary, we should consider other partitions breaking more than one translational invariance to study the entanglement signature of topological boundary states. For this purpose, we   show the physical energy spectrum with open boundary condition in $x$ and $y$ directions in Fig.~\ref{onedpart}(b). We find the parameter region  between $k_{z}=\pm\arccos (\sqrt{2} m) $ in Fig.~\ref{onedpart}(b) supports hinge arcs with four-fold degeneracy appearing in the physical hinges of the HOWSM system. Since hinge arcs are protected by   crystalline symmetries, we should consider a $2$D partition, as demonstrated in Fig.~\ref{twodpartwave}(a),  which breaks the translational invariance of $x$ and $y$ directions, but preserves the other crystalline symmetries in order to study entanglement signature of hinge arcs. As shown in Fig.~\ref{onedpart}(d), the four-fold degenerate $1/2$ EMs are found in the parameter region between $k_{z}=\pm\arccos (\sqrt{2} m) $. We conclude that  $1/2$ EMs   serve as entanglement signature of hinge arcs.

Since both hinge arcs and Fermi arcs contribute to $1/2$ EMs, to further unambiguously identify hinge arcs and Fermi arcs via entanglement,  we need to study EWFs in the $2$D partition cases. As illustrated in the Fig.~\ref{twodpartwave}, since all partitions preserve translational invariance of $z$ direction,   we only need to plot the lattice in $x-y$ plane to demonstrate the $2$D partition. Then we divide the plane into two parts A and B, and trace over part-B in order to demonstrate EWFs of $1/2$ EMs of part-A, where the diameter of each black circle denotes the EWF amplitude. We discover that the EWFs are  localized at the 4 corners of part-A, as shown in Fig.~\ref{twodpartwave}(a). In other words, the EWFs are localized on the virtual hinges of the $2$D partition. In conclusion, this entanglement signature of hinge arcs is different with that of Fermi arcs.

\begin{figure}[htbp]
  \centering
  \includegraphics[width=8.5cm]{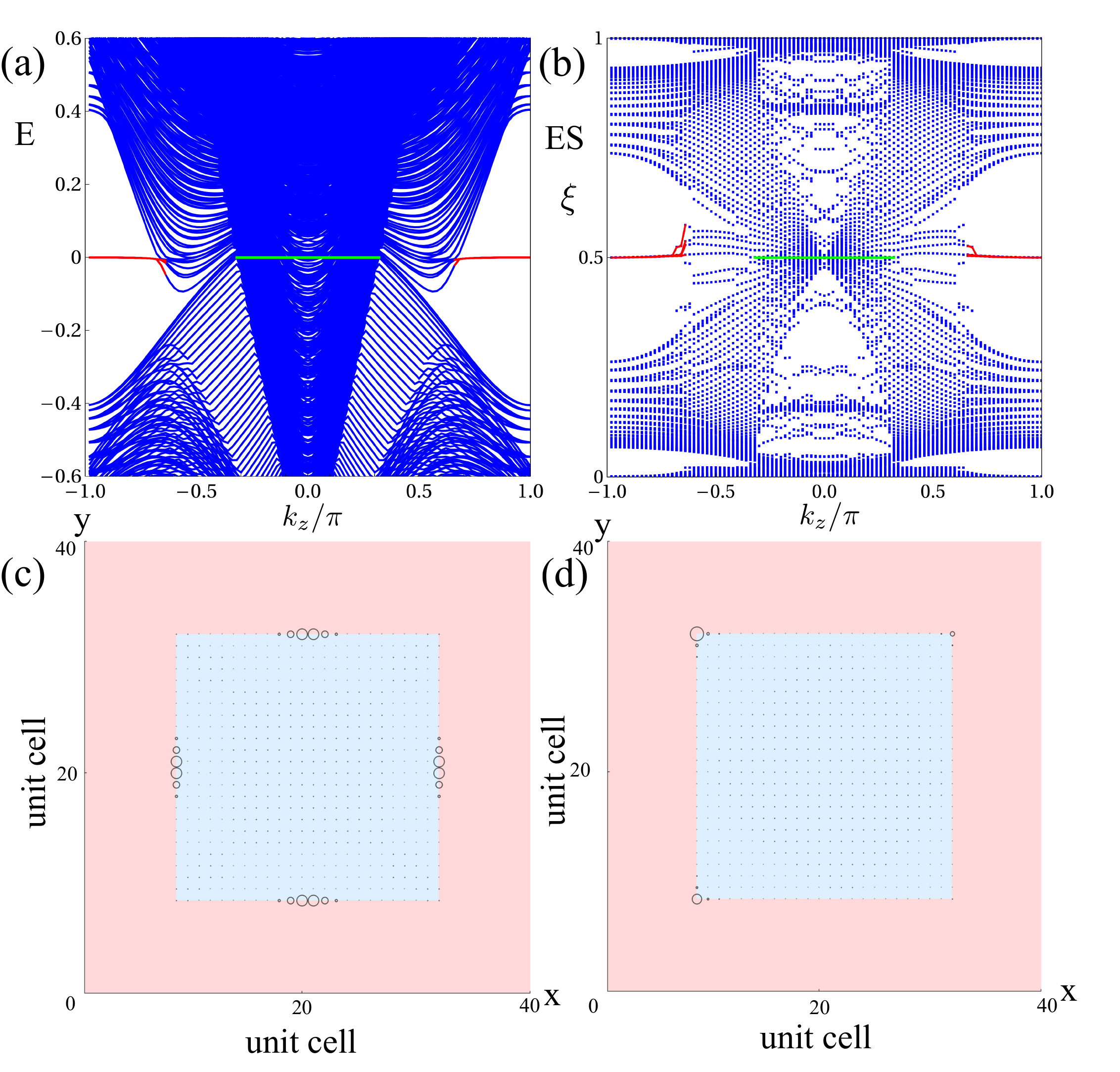}
  \caption{(a) is energy spectrum as a function of $k_{z}$  with open boundary condition along $x$ and $y$ directions, where the regions of red line and green line respectively correspond to hinge arcs and bulk gapless states. (b) is the ES as a function of $k_{z}$ with $2$D partition along both $x$ and $y$ directions, where the regions of red line and green line have $1/2$ EMs. (c) and (d) are the amplitude distribution of EWFs of $1/2$ EMs for $k_{z}=0$ and $k_{z}=\pi$ in (b), respectively. }
  \label{bulkmode}
\end{figure}

{\color{blue}\emph{Crystalline symmetry protection of $1/2$ EMs}}---
Usually, there are various crystalline symmetries exiting in higher-order topological phases, but only a part of these symmetries are needed to protect the phases. For the HOWSM model considered here, it has various crystalline symmetries, such as rotation and mirror symmetries. Below we use ES and EWFs to identify the crystalline symmetries protecting hinge arcs. In Fig.~\ref{twodpartwave} (b-c), we demonstrate two different $2$D partitions in which   the $x-y$ plane is partitioned into two parts $A$ and $B$. When tracing over $B$, based  on the Eq.~\eqref{eq_cor_matx_sym}, we find that the geometry of part $A$ determines the symmetries of ES. Specifically, part $A$ in Fig.~\ref{twodpartwave}(b) preserves two mirror symmetries $M_{x}T$ and $M_{y}T$, but breaks $C^{z}_{4}$ rotation symmetry, while the $A$ part in Fig.~\ref{twodpartwave} (c) preserves a mirror symmetry $M_{x}T$, but breaks $C^{z}_{4}$ and a mirror symmetry $M_{y}T$. Furthermore, by using numerical simulation, the ES results of these two cases are obtained as shown in Fig.~\ref{twodpartbreak}(a-b). We find the ES in Fig.~\ref{twodpartbreak}(a) still has four-fold degenerate $1/2$ EMs, while the ES in Fig.~\ref{twodpartbreak}(b) only has two-fold degenerate $1/2$ EMs. Meanwhile, compared with the partition of Fig.~\ref{twodpartwave} (a), the partition of Fig.~\ref{twodpartwave}(b) breaks four-fold rotation symmetry $C^{z}_{4}$. Therefore, we find that $C^{z}_{4}$ is not needed for protecting $1/2$ EMs. Furthermore, compared with the partition of Fig.~\ref{twodpartwave} (b), the partition of Fig.~\ref{twodpartwave} (c) breaks mirror symmetry $M_{y} T$. And the four-fold degenerate $1/2$ EMs become two-fold degeneracy. When we choose a $2$D partition breaking all crystalline symmetries, all $1/2$ EMs disappear. Meanwhile, the EWFs of these $1/2$ EMs are localized at the corner of part $A$  as shown in Fig.~\ref{twodpartwave}(b) and (c). Based on numerical results of these $2$D partitions, we find that mirror symmetry protects the $1/2$ EMs. Meanwhile, due to the entanglement signature of hinge arcs, we conclude that hinge arcs of our studying HOWSM model are protected by mirror symmetry.

{\color{blue}\emph{Entanglement contribution from  bulk gapless states}}---
For the above HOWSM model, all $1/2$ EMs can correspond to topological boundary states (either Fermi arcs or hinge arcs), so the bulk Weyl nodes do not contribute to any $1/2$ EMs. However, as discussed in Refs.~\cite{peschelReduced2009,peschelReduced2004,cheongOperatorbased2004}, even one-dimensional gapless systems without topological boundary states can contribute to $1/2$ EMs. Therefore,  in   HOWSM models with much more complex crystal structure, the bulk gapless states may potentially contribute to $1/2$ EMs, but  we can still use EWFs to distinguish them from the $1/2$ EMs corresponding to topological boundary states.

To justify our observation, we turn to the much more complex HOWSM model designed in Ref.~\cite{wangHigherOrder2020}. Due to the complex crystal structure,  this model has complex Fermi surface. In Fig.~\ref{bulkmode}(a), we present the energy spectrum with open boundary condition in $x$ and $y$ directions. We find that,  the eigenstates of zero-energy points in the red (green) line in Fig.~\ref{bulkmode}(a)  are    hinge arcs (bulk gapless states). To demonstrate the potential entanglement contribution from bulk gapless states, we apply  $2$D partitions.    In Fig.~\ref{bulkmode}(b),  ES in both red and green lines  consists of $1/2$ EMs. For the region of bulk gapless states, we choose $k_{z}=0$ as a representative point to demonstrate EWFs of $1/2$ EMs, and find the EWF is \textit{not localized} on the virtual hinges of the partition as illustrated in Fig.~\ref{bulkmode} (c). For the region of hinge arcs where we choose $k_z=\pi$, the EWF of the $1/2$ EMs is \textit{localized} (to be verified soon) on the virtual hinges of the partition as shown in Fig.~\ref{bulkmode} (d). Furthermore, due to the absence of four-fold rotation symmetry in this model, the amplitude distribution of EWFs corresponding to hinge arcs does not have four-fold rotation symmetry as illustrated in Fig.~\ref{bulkmode} (d).

To numerically verify  localization/extension of EWFs, we should carefully compare the properties of EWFs of various $1/2$ EMs in Fig.~\ref{bulkmode} (b) for different size of part $A$. From our large-size numerical simulation, we find the EWFs corresponding to hinge arcs are  \textit{localized} states, while the EWFs of $1/2$ EMs at $k_{z}=0$ in Fig.~\ref{bulkmode}(b) are  not \textit{localized} states (More  technical discussion see Appendix \ref{app_EWF}). Therefore, the $1/2$ EMs of $k_{z}=0$ in Fig.~\ref{bulkmode}(b) correspond to bulk gapless states. In this way,  we are able to use EWFs to unambiguously identify the $1/2$ EMs contributed by bulk gapless states in much complex HOWSM models.

{\color{blue}\emph{Outlook}}---We have designed various partitions with different crystalline symmetries to study the entanglement signature of Fermi arcs and hinge arcs as well as the minimal requirement of crystalline symmetries.  Our results clearly indicate that, EWFs, which have been rarely made full use of (compared to ES) in gapped systems, are  a very powerful  and necessary entanglement tool in gapless phases. Next, we may ask:  can the behavior of EWFs be measured directly by some physical quantities?     Due to the non-sparsity of correlation matrix, with the increasing size of     part $A$, we find that the time and memory consumption of numerically diagonalizing correlation matrix   increase very quickly. Then a more efficient method of obtaining ES and EWFs is very desirable. As discussed in Ref.~\cite{wangTopological2016}, the doped Weyl semimetals exhibit rich phase diagram. We may   ask whether  the entanglement signature in this paper can be used to characterize ``higher-order Weyl metals''.

The work was supported in part by Guangdong Basic and Applied Basic Research Foundation under Grant No.~2020B1515120100, NSFC Grant (No.~11847608 \& No.~12074438).


 \bibliography{apssamp}


\appendix

\widetext

\begin{figure*}[htbp]
  \centering
  \includegraphics[width=8cm]{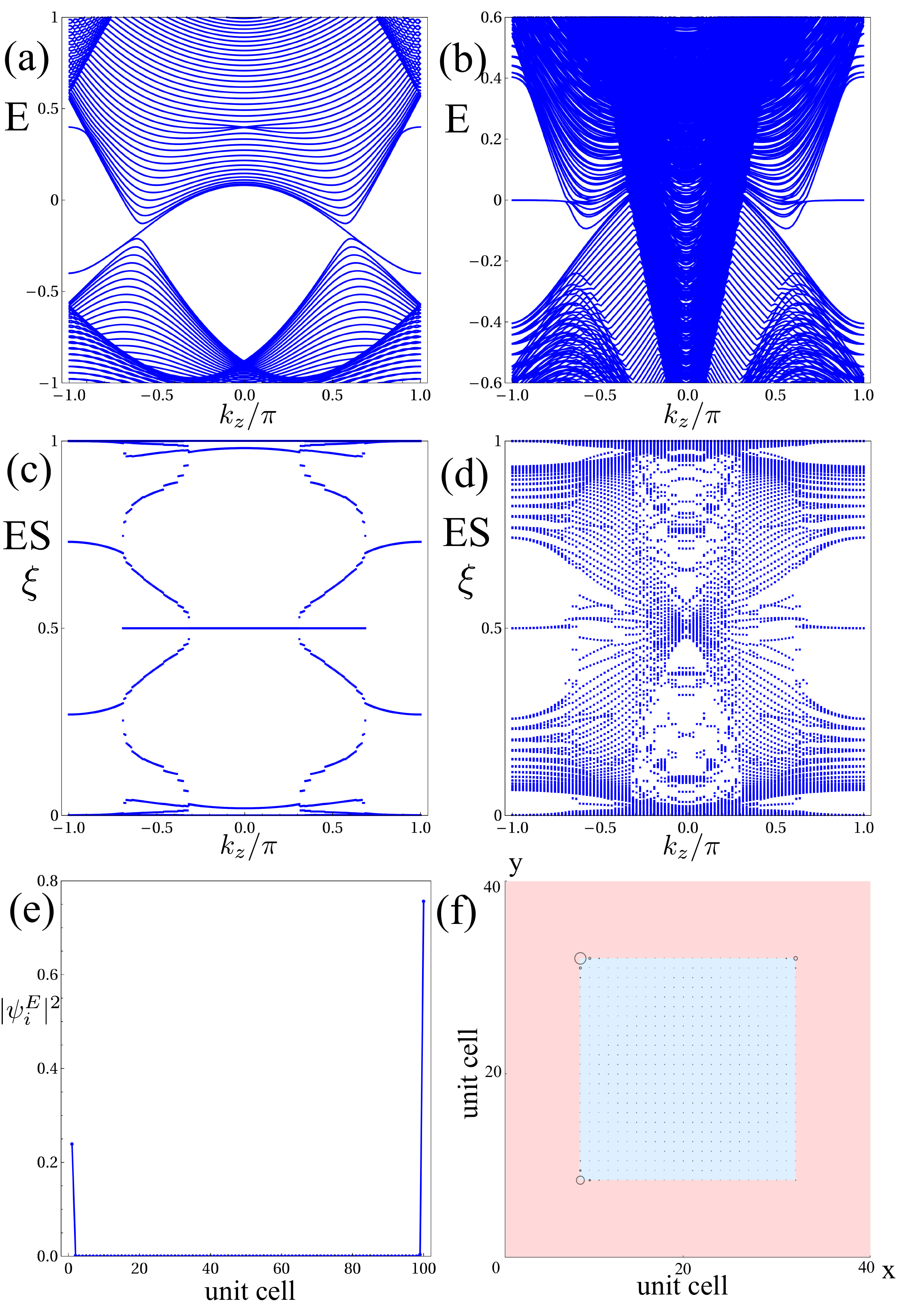}
  \caption{{\bf Energy and entanglement spectrum of the model Eq.~\eqref{secmodel} } (a) and (b) are energy spectrum of the model\eqref{secmodel} with open boundary condition respectively along $x$ direction and along $x$ and $y$ directions. (c) and (d) are the ES of the model\eqref{secmodel} with   respectively along $x$ direction (i.e., $1$D partition) and   along $x$ and $y$ directions (i.e., $2$D partition). (e) shows the amplitude distribution of EWF of $1/2$ EMs with $k_{y}=\pi$ and $k_{z}=0$ in (c). (f) shows the amplitude distribution of EWF of $1/2$ EMs with $k_{z}=\pi$ in (d). }
  \label{secondmod}
\end{figure*}

\section{The entanglement signature of HOWSM system is model-independent.}\label{app_secdhowsm}
The HOWSM phase can appear in   lattice systems with  different crystalline symmetries.   Especially, the model Hamiltonian of HOWSM in Ref.~\cite{wangHigherOrder2020} is defined on tetragonal lattices and given as:
\begin{equation}
  H(\bm{k})=
  \begin{bmatrix}
    0 & (1+e^{-i k_{z}})(t_{1}+t_{2}e^{-i(k_{x}+k_{y})}) & \gamma + \lambda e^{-i k_{x}} & \gamma + \lambda e^{-i k_{y}} \\
    & 0 & e^{i k_{z}}(\gamma + \lambda e^{i k_{y}}) & \gamma + \lambda e^{i k_{x}} \\
    & & 0 & (1+e^{-i k_{z}})(t_{1}+t_{2}e^{-i(-k_{x}+k_{y})}) \\
    & h.c. & & 0 \\
  \end{bmatrix},
  \label{secmodel}
\end{equation}
where $t_{1,2}$, $\gamma$ and $\lambda$ are the parameters. Unlike the model in Ref.~\cite{ghorashiHigherOrder2020}, this model has a four-fold screw symmetry $S_{4z}$ and two-fold rotation symmetries $C_{2x}$ and $C_{2y}$.

For this model, we also consider $1$D and $2$D partitions to study its ES. As shown in Fig.~\ref{secondmod}(c) and (d), we find both Fermi arcs and hinge arcs contribute to $1/2$ EMs in this model, where these two kinds of $1/2$ EMs are localized on the virtual surfaces and hinges of part-A (see Fig.~\ref{twodpartwave}(a) in the main text), respectively. Then, we use ES and EWF as entanglement signature to characterize hinge arc and Fermi arc, and this entanglement signature is model-independent for the HOWSM system. Remarkably, besides the $1/2$ EMs from the topological boundary states, we also find the $1/2$ EMs from the bulk gapless modes as shown in Fig.~\ref{secondmod} (d).

\begin{figure*}[htbp]
  \centering
  \includegraphics[width=10cm]{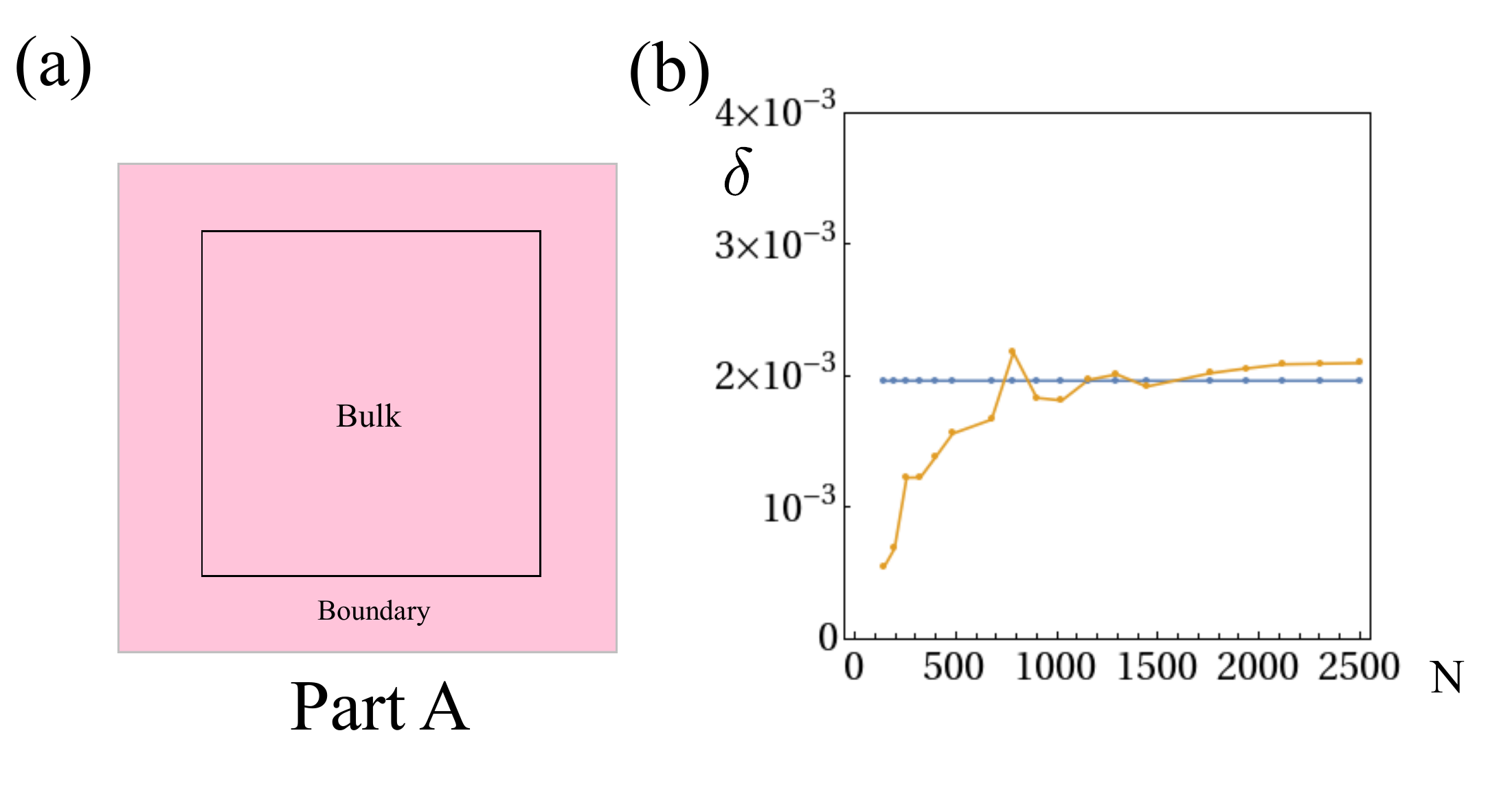}
  \caption{ (a) is the A part  of $2$D partition, and we divide A into bulk and boundary parts. The bulk amplitude of EWF of two kinds of $1/2$ EMs in Fig.~\ref{secmodel}(d) vary with the size of $2$D partition. The blue line represents $1/2$ EMs of $k_{z}=0$ , while the orange line represent $1/2$ EMs of $k_{z}=\pi$. Here $N$ is the number of unit cell of A, we fix the size of system. }
  \label{twoewf}
\end{figure*}

\section{The properties of EWF of various $1/2$ EMs in Fig.~\ref{bulkmode} (b) } \label{app_EWF}
To distinguish  various $1/2$ EMs in Fig.~\ref{bulkmode} (b), we study the properties of their EWF with different size of part-A.   As shown in Fig.~\ref{twoewf}(a), we divide part-A into the bulk and boundary parts, while the thickness of boundary part is fixed. Meanwhile, we sum the amplitude of EWF in the bulk part as a quantity $\delta =\sum_{i\in bulk} |\psi^{E}_{i}| $ named as bulk amplitude. Next, we study how $\delta$ is changed when the size of part-A is changed. First, we consider the properties of $1/2$ EMs of hinge arcs. As shown in Fig.~\ref{twoewf} (b), when we increase the size of part-A, we find the bulk amplitude of its EWF is invariant, which is numerically true for the large range of  $N$.  In contrast, for the $1/2$ EMs of $k_{z}=0$, the bulk amplitude of its EWF increases when we  increase the size of part-A.

These behaviors of EWFs can be understood as follows. For the states localized inside the boundary (finite thickness) of part-A, when its localization length is invariant and the thickness of boundary part is fixed in Fig.~\ref{twoewf}(a), its amplitude in the boundary part of part-A is invariant when we change the size of part-A. In other words, the bulk amplitude of localized states is invariant. Then, the EWFs of $1/2$ EMs associated with hinge arcs are boundary states. For the extended states, the bulk amplitude depends on the proportion of bulk part for the whole system. Meanwhile, since the thickness of the boundary part is fixed, when we increase the size of A, the proportion of bulk part becomes more and more bigger. Then, the bulk amplitude of extended states become more and more bigger. So the EWFs of the $1/2$ EMs of $k_{z}=0$ in Fig.~\ref{bulkmode}(b) is not boundary states of the entanglement Hamiltonian. In our numerical practice, the single-particle correlation matrix is not a sparse matrix, then when the size of part $A$ increases, the complexity of numerically diagonalizing correlation matrix increases very quickly.  Due to the relation between single-particle correlation matrix and entanglement Hamiltonian in free-fermion system, the entanglement Hamiltonian is not a local Hamiltonian, which needs much computational cost.

\providecommand{\noopsort}[1]{}\providecommand{\singleletter}[1]{#1}%
\end{document}